\documentclass[journal]{IEEEtran}
\ifCLASSINFOpdf
\else
\fi
\usepackage{multirow}
\usepackage{amsfonts}
\usepackage{rotating}
\usepackage{enumitem}
\usepackage{amsbsy}
\usepackage{cite}
\usepackage{amstext}
\newcommand{\eq}[1]{(\ref{#1})} 

\usepackage{xcolor}
\definecolor{cred}{HTML}{B8221E}

\definecolor{cblue}{HTML}{1965B0}

\usepackage{url}

\usepackage{breakurl}

\begin{document}
%
\title{Multitaper-mel spectrograms for keyword spotting}
%
%

\author{Douglas~Baptista~de~Souza,
        Khaled~Jamal~Bakri,
        Fernanda~Ferreira,
        and~Juliana~Inacio
\thanks{D. Baptista de Souza was with SiDi, Rua Agua\c{c}\'u 171, Pr\'edio Jacarand\'a, Loteamento Alphaville, Campinas, 13098321, Campinas, SP, Brazil, e-mail: dougdabaso@gmail.com. K. J. Bakri, F. Ferreira, and J. Inacio are with SiDi, e-mail: \{k.bakri, fernanda.f, j.camilo\}@sidi.org.br}
\thanks{Manuscript received June XX, 2022; accepted xxxxxx xx, 2022. The associate editor coordinating the review of this
manuscript was XXX XXXXXXX.}

\thanks{This letter has supplementary downloadable material available at http://ieeexplore.ieee.org, provided by the author.}}

%
%

\markboth{IEEE Signal processing letters,~Vol.~XX, No.~X, XX~XX}%
{Shell \MakeLowercase{\textit{et al.}}: Bare Demo of IEEEtran.cls for IEEE Journals}
%



\maketitle

\begin{abstract}
Keyword spotting (KWS) is one of the speech recognition tasks most sensitive to the quality of the feature representation. However, the research on KWS has traditionally focused on new model topologies, putting little emphasis on other aspects like feature extraction. This paper investigates the use of the multitaper technique to create improved features for KWS. The experimental study is carried out for different test scenarios, windows and parameters, datasets, and neural networks commonly used in embedded KWS applications. Experiment results confirm the advantages of using the proposed improved features.
\end{abstract}



%
\IEEEpeerreviewmaketitle

\section{Introduction}
%
%
%
%

\IEEEPARstart{K}{eyword} spotting (KWS) has grown in popularity in the past years, mostly due to the rise of voice-controlled systems \cite{dnn_kws_google_2014, cnn_kws_google_2015}. Keyword spotting is a subtask of automatic speech recognition (ASR), which has been traditionally carried out by full ASR systems. One difference between KWS and full ASR systems is that the former is focused on identifying a small set of pre-defined keywords. Thus, from the viewpoint of statistical classification, KWS is a simpler task than that of a full ASR system. However, KWS frameworks should run on small devices with low computational complexity \cite{kwslowcomptpower, kwsdedicatedchip}.

Addressing the tradeoff between complexity and detection performance has been one of the main challenges of KWS systems. Early solutions based on Hidden Markov Models (HMM) required modeling the keywords at phone level and employed Viterbi decoding \cite{90sHMMKWS2, iterative_posterior_viterbi}. The research on KWS migrated to deep neural networks (DNNs) \cite{early_dnn_kws}, first with hybrid HMM-DNN systems 
associating HMM phone modeling with DNN blocks \cite{earlyhmmdnn}, followed by convolutional neural networks (CNNs) for classifying audio spectrograms \cite{cnn_kws_google_2015}. Today, CNNs combining recurrent and residual blocks can achieve good KWS performances for relatively small model sizes. Examples are the convolutional residual neural networks (CNN-ResNets) and variations \cite{google_icassp18, tcresnet, tcresnet2d}, and the convolutional recurrent neural networks (CRNNs) \cite{crnn_kws, querybyexample}. Recurrent architectures, however, tend to be more difficult to tune than residual networks.

 
Other topics important to KWS such as datasets, data augmentation, and feature selection have not gained as much attention as model architectures. Nevertheless, some works in the literature have addressed these less popular topics. For example, the Google Speech Commands dataset \cite{google_scmd} has been used for KWS performance benchmarking since its release. Some data-augmentation techniques for KWS have also been proposed, like multi-style training \cite{multistyletraining} and SpecAugment \cite{specaugmentation}. Multi-style training consists in generating artificial training samples via synthetic noise mixing to simulate different evaluation scenarios. Conversely, SpecAugment works directly in the feature domain by augmenting spectrograms that have already been computed from the signals. These data-augmentation methods have been proposed to be used in conjunction with traditional mel spectral features, which exhibit some known limitations. The problem of feature extraction for KWS has been addressed in the past by front-ends based on HMM-DNN models for automatic feature learning \cite{asru_hmmdnn, icassp_td_hw_hmmdnn}. These systems could achieve improved KWS performance when compared to parametric feature extraction. However, these models were large in size and difficult to be deployed on small devices. Also, as it is common to other automatic feature extractors in audio recognition \cite{automaticfeatlearnasc}, the approaches \cite{asru_hmmdnn} and \cite{icassp_td_hw_hmmdnn} required large amounts of training data, or running extensive multi-stage training procedures \cite{asru_hmmdnn}.

Although automatic feature learning for KWS is a promising field of research, many KWS solutions still rely on parametric feature extraction \cite{crnn_kws, google_icassp18, tcresnet, tcresnet2d}. In general, KWS models are deployed on low-power chips \cite{kwsdedicatedchip, kwshelloedge}, that could not support a DNN model for feature learning. Hence, the research on parametric feature selection is still relevant to KWS. Despite some recent works on optimized features like \cite{binary_feature_kws}, mel spectrograms and mel-frequency cepstral coefficients (MFCCs) are still the most used features for KWS, regardless of their limitations \cite{mfcc_inbook, optimizingmfccs, multitapermfcc2010, mfcc_taslp}. The fact that state-of-the-art KWS approaches such as \cite{tcresnet} still use the same features as early KWS systems (e.g., \cite{earlykwsmfcc}), suggests that the research on parametric feature extraction methods for KWS has not been fully explored yet.

In this letter, we employ multitaper-mel spectrograms as KWS features. The idea of multitapering is to compute several spectra from the same signal frame with orthonormal windows (tapers), and combine them to obtain improved estimates \cite{multitaper_timefreq_stationary, manolakis}. The multitaper method has been proposed long ago \cite{thomson_multitaper}, but it has gained little attention in the past since it was time-consuming. However, this problem has been alleviated in recent years given the computational power of microprocessors \cite{kwsdedicatedchip}. In fact, multitapering has been employed with success in speaker detection applications \cite{multipmfccspeakerverify3, mfcc_speaker_verify, multipmfccspeakerverify2}. We investigate the use of multitaper features in KWS for different windows, datasets, architectures, and background noises. The advantages of using the proposed KWS features are highlighted by the findings of this paper. These are the following: 1) multitaper-mel spectrograms outperform the baseline features in terms of KWS performance in almost all cases, 2) average inference times grow approximately linearly for the simulated cases, as suggested by the multitaper expressions.

\section{Background elements} \label{ref_backgroundelements}

\subsection{The multitaper technique} \label{ref_multitaper} 

Let $x(n)$ be a discrete-time signal with $M$ samples, and $w(n)$ a window function of size $N$. To characterize how the power spectrum of $x(n)$ varies in time, one can employ $w(n)$ and the short-time Fourier transform (STFT) to compute
\begin{equation} \label{spectrogram_of_x}
\widehat{S}_{w}(\tau,f) = \Bigg| \sum_{n=0}^{N-1} x(n + H\tau)w(n)\text{e}^{- j 2 \pi n f/N} \Bigg|^2 
\end{equation}
which is the spectrogram of $x(n)$ evaluated at time frame $\tau$ with window $w(n)$ \cite{muller_book}. In \eq{spectrogram_of_x}, $f$ is a discrete frequency variable, $H$ controls the number of samples at which $w(n)$ hops over $x(n)$, and $j=\sqrt{-1}$. If we extend the computation of \eq{spectrogram_of_x} over $\tau = 0,...,T$ time frames, we get the matrix
\begin{equation} \label{matrix_spec}
\widehat{\mathbf{S}} = [\widehat{S}_{w}(0,f),...,\widehat{S}_{w}(T,f)]
\end{equation}
where $\widehat{\mathbf{S}} \in \mathbb{R}^{N \times T}$ and $T$ is the maximal frame index at which the signal extent is not extrapolated in time, given the chosen values of $M$ and $N$. To understand the advantages of the multitaper method, one has to address the computation of the spectrogram from a statistical viewpoint. More specifically, the element of \eq{matrix_spec} at position $\tau$ can be seen as an estimate of the true power spectrum  $S(\tau,f)$ of $x(n)$ at that time frame. Thus, the spectrogram can be evaluated by measures such as bias and variance. Classical window like Hann (Hanning) can reduce the bias of the spectrogram estimation, but not the variance \cite{mfcc_taslp}. The multitaper approach allows for reducing the variance in the estimation of $S(\tau,f)$ without sacrificing the bias \cite{manolakis, multitaper_timefreq_stationary}. The multitaper estimate of $S(\tau,f)$ can be computed as
\begin{equation} \label{multitaper_est}
\widehat{S}_{w}^{(K)}(\tau,f) = \sum_{k=0}^{K-1} \lambda_k \Bigg| \sum_{n=0}^{N-1} x(n + H\tau)w_{k}(n)\text{e}^{- j 2 \pi n f/N} \Bigg|^2
\end{equation}
%

where $w_k(n)$ comes from a family of $K$ orthonormal windows (or tapers) \cite{thomson_multitaper}. In \eq{multitaper_est}, the weight of the $k^{\text{th}}$ taper is given by $\lambda_k$. We discuss the choice of these windows in Section \ref{ref_multiwindow}. Next, we present how the multitaper method can be employed to compute improved mel spectrograms.
\vspace{-0.3cm}
\subsection{Multitaper-mel spectrograms} \label{ref_multitapermel} 


Let us represent $S(\tau,f)$, the true power spectrum at frame $\tau$, in vector notation as $\mathbf{s}_\tau = [S(\tau,0),...,S(\tau,N-1)]^{\text{T}}$.
%
%
With $\mathbf{s}_\tau$ at hand, the corresponding mel spectrum at frame $\tau$ is
\begin{equation} \label{mfcc_as_vector}
\mathbf{s}^{\text{mel}}_{\tau} = \frac{1}{N_{\text{m}}} \mathbf{W}^{\text{H}}\log(\mathbf{M}_{\text{m}} \mathbf{s}_\tau)
\end{equation}
where $N_{\text{m}}$ is the number of mel filter banks considered for the operation, $\mathbf{M}_{\text{m}} \in \mathbb{R}^{N_{\text{m}} \times N}$ is the filter-bank matrix contaning $N_{\text{m}}$ filter channels, and $\mathbf{W} \in \mathbb{R}^{N \times N}$ is the DFT matrix given by $\{\mathbf{W}\}_{a,b} = \text{e}^{-2 \pi j (a-1)(b-1)/N}$. From \eq{mfcc_as_vector}, it can be seen that the stochasticity in the computation of the mel spectrum comes from $\mathbf{s}_{\tau}$ ($\mathbf{W}$ and $\mathbf{M}_{\text{m}}$ are deterministic operators) \cite{mfcc_taslp}. Since in practice we only have access to estimates of $\mathbf{s}_{\tau}$ [e.g., by means of  \eq{spectrogram_of_x} or \eq{multitaper_est}], the quality of the estimator of $\mathbf{s}_{\tau}$ affects the estimated mel spectrum. The idea behind multitapering is that, by employing a low-variance method to estimate $\mathbf{s}_{\tau}$, one could improve the statistical properties of the estimated mel spectra. Thus, the multitaper estimator of $\mathbf{s}^{\text{mel}}_{\tau}$ can be derived by using \eq{multitaper_est} to estimate the true spectrum $\mathbf{s}_{\tau}$ in \eq{mfcc_as_vector}, i.e.,
\begin{equation} \label{multimel_as_vector}
\hat{\mathbf{s}}^{\text{mel}}_{K,\tau} = \frac{1}{N_{\text{m}}} \mathbf{W}^{\text{H}}\log\Big[\mathbf{M}_{\text{m}} \widehat{S}_{w}^{(K)}(\tau,\cdot) \Big]
\end{equation}
where $\widehat{S}_{w}^{(K)}(\tau,\cdot)$ is the estimate \eq{multitaper_est} at frame $\tau$. The multitaper-mel spectrogram is obtained by extending the computation of \eq{multimel_as_vector} over all time frames (i.e., $\tau=0,...T$), which gives the desired (2D) feature. Ahead, we discuss about the possible window functions to employ in the multitaper computation.
\vspace{-0.2cm}
\section{Multitaper windows} \label{ref_multiwindow}

The choice of the windows to use in \eq{multitaper_est} is often made in the light of a particular application. Discrete prolate spheroidal sequences (DPSS) is the common choice in problems involving stationary and smooth spectrum estimation \cite{manolakis}. However, DPSS tend to perform poorly when estimating non-smooth (peaky) spectra \cite{swceoriginal}, which is often the case for speech. Here, we explore two families of windows that have shown to outperform traditional ones (like DPSS) in nonstationary and speech-related settings \cite{manolakis, mfcc_taslp}. These are the Hermite and the sine-weighted cepstrum estimator (SWCE) windows.  
\vspace{-0.3cm}
\subsection{Hermite window functions} \label{ref_hermitetapers}
%

Hermite polynomials are known for their ability to characterize changes in instantaneous frequency \cite{multitaper_timefreq_stationary}.  The first $K$ Hermite window functions can be obtained as follows:
\begin{equation} \label{hermite_function}
w_{k}(n) = \text{e}^{-n^2/2} H_{k}(n) \big / \sqrt{\pi^{1/2}2^{k}k!}
\end{equation}
where $H_k(n)$ are Hermite polynomials that follow the recursion $H_{k}(n)=2n H_{k-1}(n) - 2(k-2)H_{k-2}(n)$ for $k\geq 2$ and the initialization $H_{0}(n)=1$ and $H_{1}(n)=2n$ \cite{surrogate_stationary}. Hermite multitaper spectrograms are computed by using \eq{hermite_function} in \eq{multitaper_est}, with taper weight values commonly chosen as $\lambda_k=1/K$ \cite{surrogate_stationary, icassp_2012_surro1, icassp_2014_stattest, imp_surro_test_spl}.
\vspace{-0.6cm}
\subsection{SWCE window functions} \label{ref_swcetapers}


The SWCE tapers have outperformed other windows both in terms of computation time and statistical properties (i.e., mean square error - MSE) \cite{swceoriginal}. These tapers are computed by
\begin{equation} \label{swce_function}
w_{k}(n) =\alpha \sin[\pi n k / (N+1)]
\end{equation}
where $\alpha = \sqrt{2}/\sqrt{N+1}$ is a multiplying constant. The SWCE multitaper spectrograms are obtained by employing \eq{swce_function} to calculate \eq{multitaper_est}, as well as the weights
\begin{equation} \label{swce_weights}
\lambda_{k} = \cos[\pi  k G/(N) + \beta] \Big / \sum_{k} {\{\cos[\pi k  G/(N) ] + \beta\}}
\vspace{-0.15cm}
\end{equation}
with $G=\lfloor N/K \rfloor$ and $\beta=1$. These values have shown to provide good estimates in terms of MSE for short speech segments \cite{swceoriginal, mfcc_taslp}. To increase the contrast of the generated time-frequency (TF) images, we carried out an empirical study on different values for the SWCE parameters. Although the theoretical statistical properties of the SWCE tapers may not hold if their parameters are changed, we observed that, in general, we can get sharper TF images if the SWCE tapers are computed with i) the weights $\lambda_{k}^{2p}$ raised to a given power $p$ of two (here, $p=4$), ii) the multiplying constant in \eq{swce_function} set to $\alpha = K\sqrt{2}/\sqrt{N+1}$, and iii) $\beta=0.5$ in \eq{swce_weights}. The windows obtained by considering i) to iii) are named ``SWCE modified''.

\begin{figure}
\centering
\includegraphics[width=3.5in,height=3in]{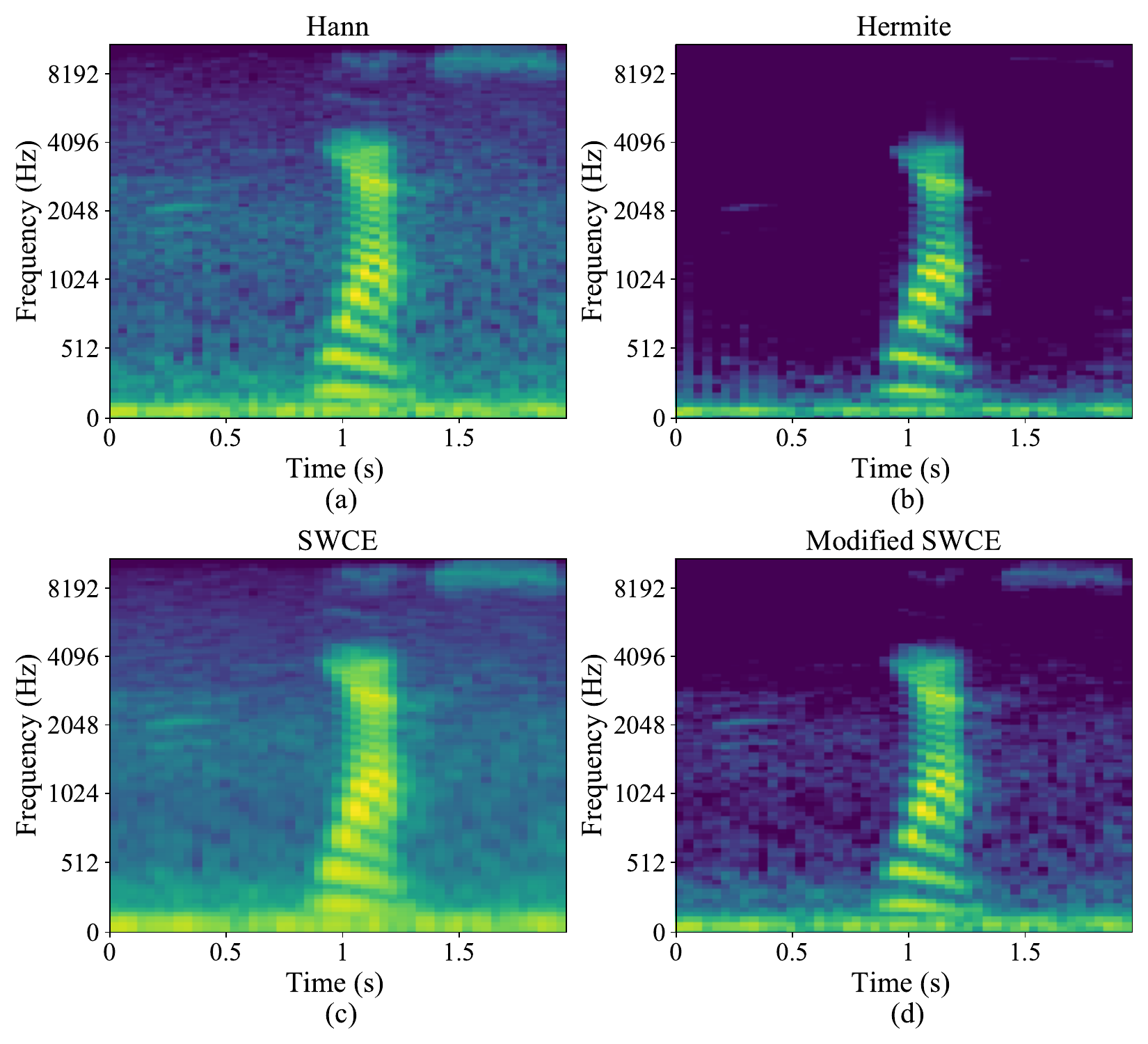}
\caption{Regular spectrogram for the default Hann window (a), and multitaper ones with (b) Hermite, (c) SWCE, and (d) modified SWCE tapers.} 
\label{fig:example_use_mel_specs}
\vspace{-0.5cm}
\end{figure}

Multitaper-mel spectrograms using $K=5$ and the Hermite, SWCE, and SWCE modified tapers are shown in Fig. \ref{fig:example_use_mel_specs}(b) to (d), and a mel spectrogram using a Hann window in Fig. \ref{fig:example_use_mel_specs}(a). The spectrograms are for one audio of the keyword ``yes'' from the Google Speech Commands dataset, which has been mixed to background noise from the Detection and Classification of Acoustic Scenes and Events (DCASE) dataset with 5 dB SNR  (see Section \ref{exp_design}) \cite{dcaseoriginal}. The STFT is computed using Setup A in Table \ref{tab:setups_config}. Notice that cleaner or smother TF images are created via multitapering, and that the modified SWCE method give sharper features than the original one. Ahead, we present the study to evaluate the proposed features. 
\vspace{-0.3cm}
\section{Experimental study} \label{ref_sectionexperiments}
\vspace{-0.05cm}
\subsection{Chosen KWS architectures}  \label{ref_kws}

In this study, aiming to probe models that could be deployed on low-power devices, we have followed the criterion of \cite{cnn_kws_google_2015} and \cite{crnn_kws} to consider KWS architectures with less than 250k parameters. The selected networks are the CNN of \cite{cnn_kws_google_2015} and the temporal convolutional residual network (TC-ResNet) of \cite{tcresnet}. The former is a CNN proposed by Google/Tensorflow research teams, and made available in \cite{tensorflowtutorial} as a tutorial on how to build a KWS system. Although \cite{cnn_kws_google_2015} is a small CNN in comparison to other speech recognition models, it has more than 250k parameters originally. To meet the model size requirement, we changed its last dense layer from 128 to 32 neurons. Also, we removed the dropouts layers in the original network to improve the KWS performance when training with fewer epochs (see Section \ref{exp_perf}). With these changes, the number of parameters of the CNN dropped from about 1.6M to less than 220k. Here, this CNN model is referred to as Tiny-CNN. The TC-ResNet of \cite{tcresnet}, is a residual neural network that treats the input TF images as pools of time-varying frequency channels, which are processed by 1D convolution blocks. By adopting this strategy, the authors in \cite{tcresnet} created a model with only 66k parameters. Block diagrams of the Tiny-CNN and TC-ResNet models are provided in the supplementary material of this letter.
\vspace{-0.15cm}
\subsection{Datasets} \label{ref_onthechosendb}
%

The experiments have been carried out using the Google Speech Commands v2 and the Mozilla Spoken Digits datasets. The former contains 105,829 samples of 35 keywords with one second of duration \cite{google_scmd}. The latter is a part of the Mozilla Common Voice \cite{commonvoice:2020} database and contains 14 keywords, namely all the digits from ``Zero'' to ``Nine'', plus the commands ``Hey'', ``Yes'', ``No'', and ``Firefox'', totalling 32,726 samples. These audios have been cropped to one second (using an energy threshold), to match the length of the Google Speech Commands samples. The same sample rate of 16 kHz has been used for both datasets, and the default train, validation, and test partitions given by the data providers have been adopted.
\vspace{-0.65cm}
\subsection{Experiment design choices}  \label{exp_design}

This experiment aims to assess noisy test cases, which are the settings in which most of real-wold KWS systems run. The test samples have been mixed with background noise using SNR values of 5, 10, and 15 dB. Two types of background noise have been tested, stationary white Gaussian noise [WGN(0,1)], and environment noise from the DCASE 2020 acoustic scene dataset. This dataset contains audio clips from different locations around the world for specific scenarios, like Park, Shopping mall, and Street traffic, for example \cite{dcaseoriginal}. 

    The multitaper-mel spectrograms have been computed using $K = 3, 5, 7, 10$. The multitaper cases are compared against baseline mel spectrograms computed using five classical windows, namely Hann, Hamming, Bartlett, Boxcar, and Kaiser with parameter $\alpha=8.168$. For more details about these windows and parameter choice, see \cite{lathiwindows}. We remark that Hann is commonly used in deep learning frameworks for audio/speech, being the default window in the Tensorflow implementation of the STFT \cite{tensorflowtutorial}. To compute mel spectrograms, one has to specify the hop size $H$, the frame/window length $N$, the lower and upper frequency bounds to compute the DFT ($f_{\text{min}})$ and $f_{\text{max}})$, and the number $N_{\text{m}}$ of mel filter banks. Considering common values from the literature, we adopted the five setups shown in Table \ref{tab:setups_config} for the aforementioned parameters.
\vspace{-0.35cm}
\subsection{Test results} \label{exp_perf}
\begin{table}
\begin{center}
\caption{Setups for mel-spectrogram parameters $H$, $N$, $f_{\text{min}}$, $f_{\text{max}}$, $N_{\text{m}}$.}
\footnotesize
\begin{tabular}{|c|c|c|c|c|c|}
\hline
Setup & Hop size  & Frame len.  &  Min. freq.       & Max. freq.  & N. mel  \\ 
      & $H$       & $N$           &  $f_{\text{min}}$ & $f_{\text{max}}$ & $N_{\text{m}}$   \\ \hline 
      
A     & 320   & 640  &  10   & 4000  & 100   \\ \hline
B     & 320   & 320  &  10   & 4000  & 100   \\ \hline
C     & 640   & 640  &  10   & 4000  & 100   \\ \hline
D     & 320   & 640  &  10   & 4000  & 40  \\ \hline
E     & 320   & 640  &  10   & 8000  & 40  \\ \hline
\end{tabular}
\label{tab:setups_config}
\end{center}
\vspace{-0.7cm}
\end{table}
The Tiny-CNN and the TC-ResNet models have been trained considering 5 epochs. The intent has been to simulate more realistic settings of embedded KWS frameworks, which often need to be trained/enrolled quickly, on-device, and possibly in an offline manner \cite{parnami2020fewshot}. 
\begin{table}
\begin{center}
\caption{Tiny-CNN and TC-ResNet average test accuracies (over all background mixing scenarios) computed for the Google Speech Commands (GSC) and Mozilla Spoken Digits (MSD) datasets.}
\begin{tabular}{|c|c|cc|cc|} 
\hline 
\multicolumn{6}{|c|}{Tiny-CNN} \\ \hline \hline  
\multirow{2}{*}{\textbf{\begin{tabular}[c]{@{}c@{}}Window\\ type\end{tabular}}}   & \multirow{2}{*}{\textbf{\begin{tabular}[c]{@{}c@{}}Num. \\ tapers\end{tabular}}} & \multicolumn{2}{c|}{\textbf{GSC dataset}}                & \multicolumn{2}{c|}{\textbf{MSD dataset}}            \\ \cline{3-6} 
                                                                                  &                                                                                  & \multicolumn{1}{c|}{\textbf{Setup C}} & \textbf{Setup D} & \multicolumn{1}{c|}{\textbf{Setup C}} & \textbf{Setup D} \\ \hline
\textbf{\begin{tabular}[c]{@{}c@{}}Hann\end{tabular}}              & \textbf{N.A.}                                                                      & \multicolumn{1}{c|}{0.7234}           & 0.7694           & \multicolumn{1}{c|}{0.6779}           & 0.7010           \\ \hline

\textbf{\begin{tabular}[c]{@{}c@{}}Hamming\end{tabular}}              & \textbf{N.A.}                                                                      & \multicolumn{1}{c|}{0.7278}  & 0.7675 & \multicolumn{1}{c|}{0.6771}           & 0.6976  \\ \hline

\textbf{\begin{tabular}[c]{@{}c@{}}Boxcar \end{tabular}}              & \textbf{N.A.}                                                                      & \multicolumn{1}{c|}{0.7282}  & 0.7602 & \multicolumn{1}{c|}{0.6652}           & 0.6930  \\ \hline

\textbf{\begin{tabular}[c]{@{}c@{}}Bartlett\end{tabular}}              & \textbf{N.A.}                                                                      & \multicolumn{1}{c|}{0.7291}  & 0.7641 & \multicolumn{1}{c|}{0.6868}           & 0.6976  \\ \hline

\textbf{\begin{tabular}[c]{@{}c@{}}Kaiser\end{tabular}}              & \textbf{N.A.}                                                                      & \multicolumn{1}{c|}{0.7289}  & 0.7601 & \multicolumn{1}{c|}{0.6793}           & 0.6966  \\ \hline

\multirow{4}{*}{\textbf{Hermite}}                                                 & \textbf{3}                                                                       & \multicolumn{1}{c|}{0.7223}           & 0.7528           & \multicolumn{1}{c|}{0.7026}           & 0.7158           \\ \cline{2-6} 
                                                                                  & \textbf{5}                                                                       & \multicolumn{1}{c|}{0.7223}           & 0.7523           & \multicolumn{1}{c|}{0.7021}           & 0.7184           \\ \cline{2-6} 
                                                                                  & \textbf{7}                                                                       & \multicolumn{1}{c|}{0.7262}           & 0.7495           & \multicolumn{1}{c|}{0.7008}           & 0.7147           \\ \cline{2-6} 
                                                                                  & \textbf{10}                                                                      & \multicolumn{1}{c|}{0.7190}           & 0.7479           & \multicolumn{1}{c|}{0.7039}           & 0.7145           \\ \hline
\multirow{4}{*}{\textbf{\begin{tabular}[c]{@{}c@{}}SWCE \\ modified\end{tabular}}} & \textbf{3}                                                                       & \multicolumn{1}{c|}{0.7451}           & 0.7799           & \multicolumn{1}{c|}{0.6865}           & 0.7057           \\ \cline{2-6} 
                                                                                  & \textbf{5}                                                                       & \multicolumn{1}{c|}{0.7562}           & 0.7790           & \multicolumn{1}{c|}{0.7049}           & \textbf{0.7271}           \\ \cline{2-6} 
                                                                                  & \textbf{7}                                                                       & \multicolumn{1}{c|}{0.7509}           & 0.7647           & \multicolumn{1}{c|}{\textbf{0.7077}}           & 0.7231           \\ \cline{2-6} 
                                                                                  & \textbf{10}                                                                      & \multicolumn{1}{c|}{0.6951}           & 0.6975           & \multicolumn{1}{c|}{0.6682}           & 0.6737           \\ \hline
\multirow{4}{*}{\textbf{\begin{tabular}[c]{@{}c@{}}SWCE\\ original\end{tabular}}} & \textbf{3}                                                                       & \multicolumn{1}{c|}{0.7500}           & 0.7648           & \multicolumn{1}{c|}{0.6804}           & 0.6881           \\ \cline{2-6} 
                                                                                  & \textbf{5}                                                                       & \multicolumn{1}{c|}{0.7600}           & 0.7715           & \multicolumn{1}{c|}{0.6970}           & 0.6961           \\ \cline{2-6} 
                                                                                  & \textbf{7}                                                                       & \multicolumn{1}{c|}{0.7704}           & \textbf{0.7849}           & \multicolumn{1}{c|}{0.6976}           & 0.6982           \\ \cline{2-6} 
                                                                                  & \textbf{10}                                                                      & \multicolumn{1}{c|}{0.\textbf{7714}}           & 0.7755           & \multicolumn{1}{c|}{0.6929}           & 0.7026           \\ \hline \hline
                                                                                  
\multicolumn{6}{|c|}{TC-ResNet} \\ \hline \hline

\textbf{\begin{tabular}[c]{@{}c@{}}Hann\end{tabular}}              & \textbf{NA}                                                                    & \multicolumn{1}{c|}{0.6342}           & 0.6420           & \multicolumn{1}{c|}{0.6017}           & 0.6297           \\ \hline

\textbf{\begin{tabular}[c]{@{}c@{}}Hamming\end{tabular}}              & \textbf{N.A.}                                                                      & \multicolumn{1}{c|}{0.6204}  & 0.6404 & \multicolumn{1}{c|}{0.6332}           & 0.6197  \\ \hline

\textbf{\begin{tabular}[c]{@{}c@{}}Boxcar \end{tabular}}              & \textbf{N.A.}                                                                      & \multicolumn{1}{c|}{0.5969}  & 0.6487 & \multicolumn{1}{c|}{0.5631}           & 0.6279  \\ \hline

\textbf{\begin{tabular}[c]{@{}c@{}}Bartlett\end{tabular}}              & \textbf{N.A.}                                                                      & \multicolumn{1}{c|}{0.6195}  & 0.6626 & \multicolumn{1}{c|}{0.6105}           & 0.6435  \\ \hline

\textbf{\begin{tabular}[c]{@{}c@{}}Kaiser\end{tabular}}              & \textbf{N.A.}                                                                      & \multicolumn{1}{c|}{0.6083}  & 0.6577 & \multicolumn{1}{c|}{0.6335}           & 0.6384  \\ \hline

\multirow{4}{*}{\textbf{Hermite}}                                                 & \textbf{3}                                                                     & \multicolumn{1}{c|}{0.4922}           & 0.4817           & \multicolumn{1}{c|}{0.5878}           & 0.5943           \\ \cline{2-6} 
                                                                                  & \textbf{5}                                                                     & \multicolumn{1}{c|}{0.5251}           & 0.5497           & \multicolumn{1}{c|}{0.5802}           & 0.6139           \\ \cline{2-6} 
                                                                                  & \textbf{7}                                                                     & \multicolumn{1}{c|}{0.5031}           & 0.5134           & \multicolumn{1}{c|}{0.5524}           & 0.5917           \\ \cline{2-6} 
                                                                                  & \textbf{10}                                                                    & \multicolumn{1}{c|}{0.4988}           & 0.5173           & \multicolumn{1}{c|}{0.5742}           & 0.5820           \\ \hline
\multirow{4}{*}{\textbf{\begin{tabular}[c]{@{}c@{}}SWCE \\ modified\end{tabular}}} & \textbf{3}                                                                     & \multicolumn{1}{c|}{0.6202}           & 0.6489           & \multicolumn{1}{c|}{0.6163}           & 0.6361           \\ \cline{2-6} 
                                                                                  & \textbf{5}                                                                     & \multicolumn{1}{c|}{0.5629}           & 0.5960           & \multicolumn{1}{c|}{0.6207}           & 0.6539           \\ \cline{2-6} 
                                                                                  & \textbf{7}                                                                     & \multicolumn{1}{c|}{0.5497}           & 0.5584           & \multicolumn{1}{c|}{0.5938}           & 0.6001           \\ \cline{2-6} 
                                                                                  & \textbf{10}                                                                    & \multicolumn{1}{c|}{0.4489}           & 0.4121           & \multicolumn{1}{c|}{0.5012}           & 0.5239           \\ \hline
\multirow{4}{*}{\textbf{\begin{tabular}[c]{@{}c@{}}SWCE \\ original\end{tabular}}} & \textbf{3}                                                                     & \multicolumn{1}{c|}{0.6460}           & 0.6479           & \multicolumn{1}{c|}{0.6457}           & 0.6372           \\ \cline{2-6} 
                                                                                  & \textbf{5}                                                                     & \multicolumn{1}{c|}{0.6373}           & 0.6411           & \multicolumn{1}{c|}{0.6628}           & 0.6836           \\ \cline{2-6} 
                                                                                  & \textbf{7}                                                                     & \multicolumn{1}{c|}{0.6520}           & \textbf{0.6734}           & \multicolumn{1}{c|}{0.6732}           & \textbf{0.7077}           \\ \cline{2-6} 
                                                                                  & \textbf{10}                                                                    & \multicolumn{1}{c|}{\textbf{0.6593}}           & 0.6646           & \multicolumn{1}{c|}{\textbf{0.7041}}           & 0.6833           \\ \hline
                                                                                  
\end{tabular}
\vspace{-0.6cm}
\label{tab:performance_tinytc}
\end{center}
\end{table}
The selected architectures have been trained and tested three times using the datasets, setups, and testing scenarios described previously, totalling 2040 tested cases (2 models x 2 datasets x 17 window configurations x 5 STFT setups x 6 background noises = 2040). For each case, we have computed the average classification accuracy over the three test runs\footnote{Although data augmentation can help boosting the KWS performance, it has not been added to the training pipeline with the intent to isolate the effects of using the proposed features, which is the focus of the paper. The use of data augmentation with multitaper features will be considered in future works.}. The results for all cases are given in the supplementary material of this letter. Here, we summarize the test results of Setups C and D in Table \ref{tab:performance_tinytc}, which shows the average accuracies over all background noises for these setups (best cases shown in bold). Note that the baseline windows are outperformed in all cases of Table
\ref{tab:performance_tinytc}, and in the vast majority of cases in the supplementary material tables. In general, the best accuracies are for the SWCE window and its modified version. In particular, the cases SWCE modified yields the best results are for the Tiny-CNN. This network is more prone to benefit from the improved contrast of the modified SWCE features, since it employs larger convolution kernels and has three times the number of parameters of the TC-ResNet. Nevertheless, considering the overall test performance, the original SWCE features give the best results the most often. The performances obtained by the SWCE methods are in line with the good statistical properties of the SWCE windows \cite{swceoriginal}. Moreover, observe that the performance of SWCE modified tend to decrease with of $K$, notably for $K=10$. We note that parameter $\alpha$ for this window is obtained by multiplying the original parameter by $K$, which can make the SWCE modified window too sensitive to large $K$. Finally, we remark that the multitaper parameters can be considered by future frameworks dedicated to automatic learning of KWS feature representation.

\begin{figure}
\centering
\includegraphics[width=2.6in,height=1.8in]{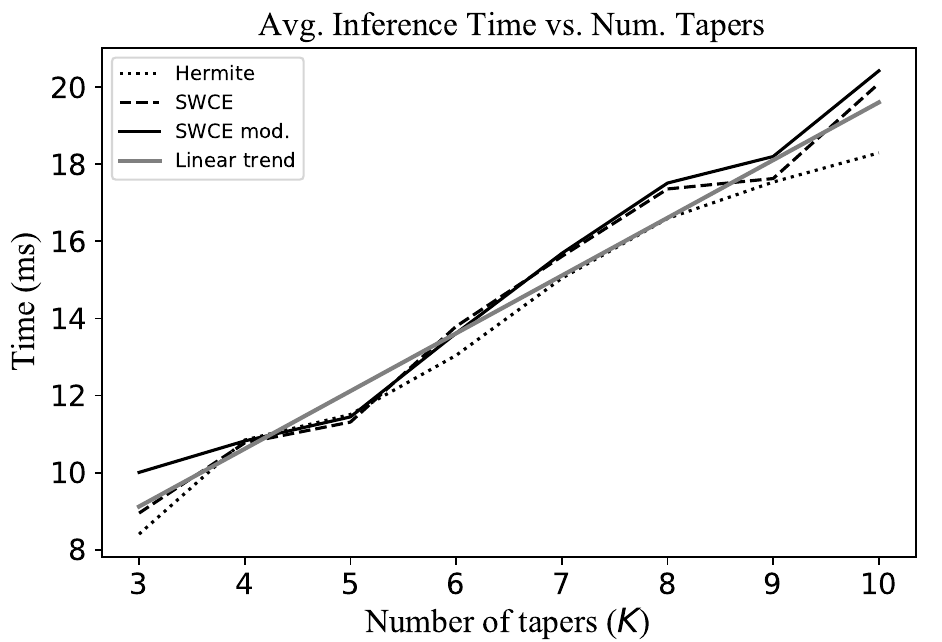}
\caption{For the TC-ResNet model, average inference time vs. $K$ by using different multitaper-mel spectrogram features. For the simulations,  ``yes''  audios samples from the Google Speech Commands dataset have been tested.} 
\label{fig:inferencetime}
\vspace{-0.5cm}
\end{figure}

To complement the study on the KWS models with multitaper features, we have also evaluated how the inference times grow with $K$ for the multitaper windows. To evaluate this scenario, we have computed the average inference time for the TC-ResNet and multitaper features using all values of $K \in [3,10]$. The measurements have been made over individual audios of the Google Speech Commands dataset belonging to the ``yes'' class, and a particular noise scenario (DCASE and 5 dB SNR). The simulations have been performed on a single CPU Intel i7-8565U 1.80GHz. A linear trend fitted to the average of the three curves is shown in Fig. \ref{fig:inferencetime}. Notice that the average inference times grow approximately linearly with $K$, as suggested by the multitaper expressions. These results favor the idea that the multitaper features could be implemented in practical devices, yielding improved classification performances and inference times that grow linearly with $K$.
\vspace{-0.1cm}
\section{Conclusions} \label{ref_secconclusion}

In this paper, we investigated the use of multitaper-mel spectrograms as features for keyword spotting (KWS). Different windows were considered to compute multitaper features. For the experimental study, KWS models with small footprint were selected. Keyword recognition performance and inference time were assessed in different test scenarios. When comparing the new feature extraction method to the baseline case, simulation results showed the classification performances could be improved in almost all cases, while observing the desired linear increase in inference time as the number of tapers grows.
\vspace{-0.2cm}

\subsection*{Acknowledgements}
The work presented in this paper ``Multitaper-mel spectrograms for keyword spotting'', D. Baptista de Souza et al., 2022, has been developed as part of a project between SiDi, financed by Samsung Eletr\^onica da Amazonia Ltda., under the auspices of the Brazilian Federal Law of Informatics no. 8248/91.

\ifCLASSOPTIONcaptionsoff
  \newpage
\fi



%

\bibliographystyle{IEEEbib}
\bibliography{biblio}

\begin{thebibliography}{10}

\bibitem{dnn_kws_google_2014}
G.~Chen, C.~Parada, and G.~Heigold,
\newblock ``Small-footprint keyword spotting using deep neural networks,''
\newblock in {\em Proc. IEEE Int. Conf. Acoust., Speech, Signal Process. (ICASSP)}, Florence, Italy, May 2014, pp. 4087--4091.

\bibitem{cnn_kws_google_2015}
T.~N. Sainath and C.~Parada,
\newblock ``Convolutional neural networks for small-footprint keyword spotting,''
\newblock in {\em Proc. Int. Speech Commun. Assoc. (INTERSPEECH)}, Dresden, Germany, Sep. 2015, pp. 1478--1482.

\bibitem{kwslowcomptpower}
S.~Kodali, P.~Hansen, N.~Mulholland, P.~Whatmough, D.~Brooks, and G.~Wei,
\newblock ``Applications of deep neural networks for ultra low power iot,''
\newblock in {\em Proc. IEEE Int. Conf. Comput. Des. (ICCD)}, Boston, MA, USA, 2017, pp. 589--592.

\bibitem{kwsdedicatedchip}
{Jeremy Holleman},
\newblock ``The speed and power advantage of a purpose-built neural compute engine,'' [Online]. Available: \url{https://www.syntiant.com/post/keyword-spotting-power-comparison},
\newblock {Accessed on: May. 30, 2022}.

\bibitem{90sHMMKWS2}
R.~Rose and D.~Paul,
\newblock ``A hidden markov model based keyword recognition system,''
\newblock in {\em Proc. IEEE Int. Conf. Acoust., Speech, Signal Process. (ICASSP)}, Albuquerque, NM, USA, Apr. 1990, pp. 129--132.

\bibitem{iterative_posterior_viterbi}
M.-C. Silaghi and H.~Boulard,
\newblock ``A new keyword spotting approach based on iterative dynamic programming,''
\newblock in {\em Proc. IEEE Int. Conf. Acoust., Speech, Signal Process.(ICASSP)}, Istanbul, Turkey, Jun. 2000, pp. 1831--1834.

\bibitem{early_dnn_kws}
G.~Wang and K.~C. Sim,
\newblock ``Context dependent acoustic keyword spotting using deep neural network,''
\newblock in {\em Asia-Pacific Signal Inf. Process. Assoc. Annu. Summit Conf. (APSIPA)}, Kaohsiung, Taiwan, Nov. 2013, pp. 1--10.

\bibitem{earlyhmmdnn}
Sankaran Panchapagesan, Ming Sun, Aparna Khare, Spyros Matsoukas~Arindam Mandal, Bjorn Hoffmeister, and Shiv Vitaladevuni,
\newblock ``Multi-task learning and weighted cross-entropy for {DNN}-based keyword spotting,''
\newblock in {\em Proc. Int. Speech Commun. Assoc. (INTERSPEECH)}, San Francisco, USA, Sep. 2016, pp. 760--764.

\bibitem{google_icassp18}
R.~Tang and J.~Lin,
\newblock ``Deep residual learning for small-footprint keyword spotting,''
\newblock in {\em Proc. IEEE Int. Conf. Acoust., Speech, Signal Process. (ICASSP)}, Calgary, AB, Canada, Apr. 2018, pp. 5484--5488.

\bibitem{tcresnet}
S.~Choi, S.~Seo, B.~Shin, H.~Byun, M.~Kersner, B.~Kim, D.~Kim, and S.~Ha,
\newblock ``Temporal convolution for real-time keyword spotting on mobile devices,''
\newblock in {\em Proc. Int. Speech Commun. Assoc. (INTERSPEECH)}, Graz, Austria, Sep. 2019, pp. 3372--3376.

\bibitem{tcresnet2d}
Byeonggeun Kim, Simyung Chang, Jinkyu Lee, and Dooyong Sung,
\newblock ``Broadcasted residual learning for efficient keyword spotting,''
\newblock in {\em Proc. Int. Speech Commun. Assoc. (INTERSPEECH)}, Brno, Czech Republic, Sep. 2021, pp. 4532--4542.

\bibitem{crnn_kws}
S.~Ö. Arık, M.~Kliegl, R.~Child, J.~Hestness, A.~Gibiansky, C.~Fougner, R.~Prenger, and A.~Coates,
\newblock ``Convolutional recurrent neural networks for small-footprint keyword spotting,''
\newblock in {\em Proc. Int. Speech Commun. Assoc. (INTERSPEECH)}, Stockholm, Sweden, Aug. 2017, pp. 1606--1610.

\bibitem{querybyexample}
G.~Chen, C.~Parada, and T.~N. Sainath,
\newblock ``Query-by-example keyword spotting using long short-term memory networks,''
\newblock in {\em Proc. IEEE Int. Conf. Acoust., Speech, Signal Process. (ICASSP)}, South Brisbane, QLD, Australia, Apr. 2015, pp. 5236--5240.

\bibitem{google_scmd}
P.~{Warden},
\newblock ``Speech commands: A dataset for limited-vocabulary speech recognition,''
\newblock {\em ArXiv e-prints: 1804.03209}, p. 1–11, Apr. 2018.

\bibitem{multistyletraining}
Rohit Prabhavalkar, Raziel Alvarez, Carolina Parada, Preetum Nakkiran, and Tara~N. Sainath,
\newblock ``Automatic gain control and multi-style training for robust small-footprint keyword spotting with deep neural networks,''
\newblock in {\em Proc. IEEE Int. Conf. Acoust., Speech, Signal Process. (ICASSP)}, South Brisbane, Australia, Apr. 2015, pp. 4704--4708.

\bibitem{specaugmentation}
Daniel~S. Park, William Chan, Yu~Zhang, Chung-Cheng Chiu, Barret Zoph, Ekin~D. Cubuk, and Quoc~V. Le,
\newblock ``{SpecAugment}: A simple data augmentation method for automatic speech recognition,''
\newblock in {\em Proc. Int. Speech Commun. Assoc. (INTERSPEECH)}, Graz, Austria, Sep. 2019, pp. 2613--2617.

\bibitem{asru_hmmdnn}
Kenichi Kumatani, Sankaran Panchapagesan, Minhua Wu, Minjae Kim, Nikko Str{\"o}m, Gautam Tiwari, and Arindam Mandal,
\newblock ``Direct modeling of raw audio with {DNNS} for wake word detection,''
\newblock in {\em Proc. IEEE Automatic Speech Recognition and Understanding Workshop (ASRU)}, Okinawa, Japan, Dec. 2017, pp. 252--257.

\bibitem{icassp_td_hw_hmmdnn}
Jinxi Guoy, Kenichi Kumatani, Ming Sun, Minhua Wu, Anirudh Raju, Nikko Str{\"o}m, and Arindam Mandal,
\newblock ``Time-delayed bottleneck highway networks using a {DFT} feature for keyword spotting,''
\newblock in {\em Proc. IEEE Int. Conf. Acoust., Speech, Signal Process. (ICASSP)}, Calgary, Canada, Apr. 2018, pp. 5489--5493.

\bibitem{automaticfeatlearnasc}
Hangting Chen, Pengyuan Zhang, and Yonghong Yan,
\newblock ``An audio scene classification framework with embedded filters and a {DCT}-based temporal module,''
\newblock in {\em Proc. IEEE Int. Conf. Acoust., Speech, Signal Process. (ICASSP)}, Brighton, UK, May 2019, pp. 835--839.

\bibitem{kwshelloedge}
Yundong Zhang, Naveen Suda, Liangzhen Lai, and Vikas Chandra,
\newblock ``Hello edge: Keyword spotting on microcontrollers,''
\newblock {\em ArXiv e-prints: 1711.07128}, p. 1–14, Fev. 2018.

\bibitem{binary_feature_kws}
A.~Riviello and J.-P. David,
\newblock ``Binary speech features for keyword spotting tasks,''
\newblock in {\em Proc. Int. Speech Commun. Assoc. (INTERSPEECH)}, Graz, Austria, Sep. 2019, pp. 3460--3464.

\bibitem{mfcc_inbook}
R.~Gonzalez,
\newblock {\em Better than MFCC audio classification features. In: The era of interactive media}, pp. 291--301,
\newblock Springer, New York, NY, 2013.

\bibitem{optimizingmfccs}
V.~A. Hadoltikar, V.~R. Ratnaparkhe, and R.~Kumar,
\newblock ``Optimization of mfcc parameters for mobile phone recognition from audio recordings,''
\newblock in {\em In Proc. Int. Conf. on Electron., Commun., Aerosp. Technol. (ICECA)}, Coimbatore, India, Jun. 2019, pp. 777--780.

\bibitem{multitapermfcc2010}
J.~Sandberg, M.~Hansson-Sandsten, T.~Kinnunen, R.~Saeidi, P.~Flandrin, and P.~Borgnat,
\newblock ``Multitaper estimation of frequency-warped cepstra with application to speaker verification,''
\newblock {\em IEEE Signal Process. Lett.}, vol. 17, no. 4, pp. 343--346, Jan. 2010.

\bibitem{mfcc_taslp}
T.~Kinnunen, R.~Saeidi, F.~Sedlak, K.~A. Lee, J.~Sandberg, M.~Hansson-Sandsten, and H.~Li,
\newblock ``Low-variance multitaper mfcc features: A case study in robust speaker verification,''
\newblock {\em IEEE Trans. Audio, Speech, Lang. Process.}, vol. 20, no. 7, pp. 1990--2001, Apr. 2012.

\bibitem{earlykwsmfcc}
Y.-J. Chen, C.-H. Wu, and G.-L. Yan,
\newblock ``Utterance verification using prosodic information for mandarin telephone speech keyword spotting,''
\newblock in {\em Proc. IEEE Int. Conf. Acoust., Speech, Signal Process. (ICASSP)}, Phoenix, AZ, USA, Mar. 1999, vol.~2, pp. 697--700.

\bibitem{multitaper_timefreq_stationary}
J.~Xiao and P.~Flandrin,
\newblock ``Multitaper time-frequency reassignment for nonstationary spectrum estimation and chirp enhancement,''
\newblock {\em {IEEE} {T}rans. {S}ignal {P}rocess.}, vol. 55, no. 6, pp. 2851--2860, 2007.

\bibitem{manolakis}
D.~G. Manolakis, V.~K. Ingle, and S.~M. Kogon,
\newblock {\em {Statistical and Adaptive Signal Processing: Spectral estimation, Signal Modeling, Adaptative Filtering and Array Processing}},
\newblock London, UK: {A}rtech {H}ouse, 2005.

\bibitem{thomson_multitaper}
D.J. Thomson,
\newblock ``Spectrum estimation and harmonic analysis,''
\newblock in {\em Proc. IEEE}, Sep. 1982, vol.~70, pp. 1055--1096.

\bibitem{multipmfccspeakerverify3}
K.~V. Veena and D.~Mathew,
\newblock ``Speaker identification and verification of noisy speech using multitaper mfcc and gaussian mixture models,''
\newblock in {\em Proc. Int. Conf. Power, Instrum., Control, Comput. (PICC)}, Thrissur, India, Dec. 2015, pp. 1--4.

\bibitem{mfcc_speaker_verify}
A.~Mansouri and E.~Castillo-Guerra,
\newblock ``Multitaper {MFCC} and normalized multitaper phase-based features for speaker verification,''
\newblock {\em SN Appl. Sci.}, vol. 1, no. 290, pp. 1--18, Mar. 2019.

\bibitem{multipmfccspeakerverify2}
K.~P.~Bharath ~ and R.~Kumar,
\newblock ``Multitaper based mfcc feature extraction for robust speaker recognition system,''
\newblock in {\em Proc. Innovations Power Adv. Comput. Technol. (i-PACT)}, Vellore, India, Mar. 2019, vol.~1, pp. 1--5.

\bibitem{muller_book}
M.~M{\"u}ller,
\newblock {\em Fundamentals of Music Processing. Audio, Analysis, Algorithms, Applications},
\newblock Switzerland: Springer, 2015.

\bibitem{swceoriginal}
M.~Hansson-Sandsten and J.~Sandberg,
\newblock ``Optimal cepstrum estimation using multiple windows,''
\newblock in {\em Proc. IEEE Int. Conf. Acoust., Speech, Signal Process. (ICASSP)}, Taipei, Taiwan, Apr. 2009, pp. 3077--3088.

\bibitem{surrogate_stationary}
P.~Borgnat, P.~Flandrin, P.~Honeine, C.~Richard, and J.~Xiao,
\newblock ``Testing stationarity with surrogates: A time-frequency approach,''
\newblock {\em {IEEE} {T}rans. {S}ignal {P}rocess.}, vol. 58, no. 7, pp. 3459--3470, 2010.

\bibitem{icassp_2012_surro1}
D.~Baptista de~souza, J.~Chanussot, A.~C. Favre, and P.~Borgnat,
\newblock ``A modified time-frequency method for testing wide-sense stationarity,''
\newblock in {\em Proc. IEEE Int. Conf. Acoust., Speech, Signal Process. (ICASSP)}, Kyoto, Japan, Mar. 2012, pp. 3409--3412.

\bibitem{icassp_2014_stattest}
D.~Baptista de~souza, J.~Chanussot, A.~C. Favre, and P.~Borgnat,
\newblock ``A new nonparametric method for testing stationarity based on trend analysis in the time marginal distribution,''
\newblock in {\em Proc. IEEE Int. Conf. Acoust., Speech, Signal Process. (ICASSP)}, Florence, Italy, May 2014, pp. 320--324.

\bibitem{imp_surro_test_spl}
D.~Baptista de~souza, J.~Chanussot, A.~C. Favre, and P.~Borgnat,
\newblock ``An improved stationarity test based on surrogates,''
\newblock {\em IEEE Signal Process. Lett.}, vol. 26, no. 10, pp. 1431--1435, Oct. 2019.

\bibitem{dcaseoriginal}
D.~Stowell, D.~Giannoulis, E.~Benetos, M.~Lagrange, and M.~D. Plumbley,
\newblock ``Detection and classification of acoustic scenes and events,''
\newblock {\em IEEE Trans. Multimedia}, vol. 17, no. 10, pp. 1733--1746, May 2015.

\bibitem{tensorflowtutorial}
{Tensorflow},
\newblock ``Simple audio recognition: Recognizing keywords,'' [Online]. Available: \url{https://www.tensorflow.org/tutorials/audio/simple_audio},
\newblock {Accessed on: Jan. 30, 2022}.

\bibitem{commonvoice:2020}
R.~Ardila, M.~Branson, K.~Davis, M.~Henretty, M.~Kohler, J.~Meyer, R.~Morais, L.~Saunders, F.~M. Tyers, and G.~Weber,
\newblock ``Common voice: A massively-multilingual speech corpus,''
\newblock in {\em Proc. Conf. Lang. Resour. Eval. (LREC)}, Dresden, Germany, May 2020, pp. 4211--4215.

\bibitem{lathiwindows}
B.~P. Lathi,
\newblock {\em Linear Systems and Signals}, p. 750,
\newblock {N}ew {Y}ork, {NY}, {USA}: {O}xford {U}niversity {P}ress, 2 edition, 2004.

\bibitem{parnami2020fewshot}
Archit Parnami and Minwoo Lee,
\newblock ``Few-shot keyword spotting with prototypical networks,'' 2020.

\end{thebibliography}

%








\end{document}